
\magnification=1200
\hoffset=.2truein \voffset=-.5in
\hsize=6.5truein\vsize=8.4truein
\parindent 10pt
\parskip 1pt plus 2pt
\baselineskip=20pt

\overfullrule=0pt

{}~\vbox{\hbox{TIFR/TH/92-20}\hbox{hepth@xxx/9204046}\hbox{April,
1992}}\break

\vskip 1in

\centerline{\bf ROTATING CHARGED BLACK HOLE SOLUTION }
\centerline{\bf IN HETEROTIC STRING THEORY}

\vskip 1in

\centerline{Ashoke Sen}

\centerline{\it Tata Institute of Fundamental Research}
\centerline{\it Homi Bhabha Road, Bombay
400005, India}
\centerline{e-mail address: sen@tifrvax.bitnet}

\vskip 1in

\centerline{\bf Abstract}

We construct a solution of the classical equations of motion
arising in the low energy effective field theory for heterotic string
theory.
This solution describes a black hole in four dimensions carrying mass $M$,
charge $Q$ and angular momentum $J$.
The extremal limit of the solution is discussed.

\vfill\eject

\def\BLACK{[1]}
\def\ABS{[2]}

\def\FAWAD{[7]}
\def\DUFF{[4]}
\def\ODD{[8]}
\def\HOROWITZ{[3]}
\def\FROLOV{[9]}

\def\p{\partial}
\def\ck{{\cal K}}
\def\cm{{\cal M}}

\def\eone{(1)}
\def\etwo{(2)}
\def\ethree{(3)}
\def\ethreenew{(4)}
\def\efour{(5)}
\def\efive{(6)}
\def\esix{(7)}
\def\eseven{(8)}
\def\eeight{(9)}
\def\enine{(10)}
\def\eten{(11)}
\def\eeleven{(12)}
\def\etwelve{(13)}
\def\ethirteen{(14)}
\def\efourteen{(15)}
\def\efifteen{(16)}
\def\esixteen{(17)}
\def\eeighteen{(18)}
\def\enineteen{(19)}
\def\etwenty{(20)}
\def\etwentyfour{(21)}
\def\etwentyfive{(22)}
\def\etwentyfoura{(23)}
\def\etwentyeight{(24)}
\def\etwentyeighta{(25)}
\def\etwentynine{(26)}

It has been realized recently that the low energy effective field theory
describing string theory contains black hole (or, more generally, black
$p$-brane) solutions which can have properties which are qualitatively
different from those that appear in ordinary Einstein gravity\BLACK.
Most of these solutions are characterized by one or more charges
associated with Yang-Mills fields or the antisymmetric tensor gauge field,
and a non-trivial dilaton field.
In the absence of any charge, the solution reduces to the ordinary
Schwarzschild solution.

Rotating charge neutral black hole solutions can also be constructed in
string theory,
and are identical to the Kerr solution\ABS\ of ordinary Einstein gravity
with the dilaton taking a constant value.
Recently, rotating charged black hole solutions in these theories have
been analyzed\HOROWITZ\ in the limit of small angular momentum.
In fact, in ref.\HOROWITZ\ the authors consider a more general class of
theories than those which arise as the low energy effective action in
string theory, by allowing dilaton couplings to the Maxwell field of the
type which is not necessarily the one induced in string theory.
They, however, consider only the dilaton-graviton system, and do not
consider more general action that also includes antisymmetric tensor gauge
field.

In this paper we shall construct an exact classical solution in the low
energy effective field theory describing heterotic string theory, which
describes a black hole carrying finite amount of charge and
angular momentum.
Our solution, however, differs from that of ref.\HOROWITZ\ even in the
limit of small angular momentum since it involves the antisymmetric tensor
field in a non-trivial way.
In fact, since a rotating charged black hole also carries a magnetic
dipole moment, the antisymmetric tensor field background is induced
automatically if we take into account the coupling of the
antisymmetric tensor gauge field strength to the Chern-Simons three form
constructed from the gauge fields.

The method that we shall be using for obtaining the solution is the
twisting
procedure\DUFF-\ODD\
that generates inequivalent classical solutions starting from a given
classical solution of string theory.
In particular, in ref.\FAWAD\ it was shown how one can generate charged
blackhole solutions starting from a charge neutral solution.
With the help of the same transformations, we shall generate the rotating
charged black hole solution by starting from a rotating black hole
solution carrying no charge, i.e. the Kerr solution\ABS.

We shall first summarize the results of ref.\FAWAD\ applied to the
particular problem at hand.
We begin with the string theory effective action in four dimensions:
$$
S=-\int d^4 x\sqrt{- det~G} e^{-\Phi} (-R+{1\over
12}H_{\mu\nu\rho}H^{\mu\nu\rho} -G^{\mu\nu}\p_\mu\Phi\p_\nu\Phi
+{1\over 8}F_{\mu\nu}F^{\mu\nu})
\eqno\eone
$$
Here $G_{\mu\nu}$ is the metric, $R$ is the scalar curvature,
$F_{\mu\nu}=\p_\mu A_\nu-\p_\nu A_\mu$ is the field strength
corresponding to the Maxwell field $A_\mu$, $\Phi$ is the dilaton field,
and,
$$
H_{\mu\nu\rho} =\p_\mu B_{\nu\rho} +{~\rm cyclic~permutations}
-(\Omega_3(A))_{\mu\nu\rho}
\eqno\etwo
$$
where $B_{\mu\nu}$ is the antisymmetric tensor gauge field, and,
$$
(\Omega_3(A))_{\mu\nu\rho}={1\over 4} (A_\mu F_{\nu\rho}+ {~\rm cyclic
{}~permutations})
\eqno\ethree
$$
is the gauge Chern-Simons term.
There are several points we need to mention at this stage.
They are the followings.

\item {1.} We are considering a theory
where 6 of the 10 dimensions have been
compactified (say, to a Calabi-Yau manifold).
The massless fields arising from compactification have not been included
in the effective action.

\item
{2.} We have included only a $U(1)$ component of the full set of non-abelian
gauge fields present in the theory.
This will suffice for our purpose, since we shall look for solutions
carrying $U(1)$ charge only.

\item {3.} The metric $G_{\mu\nu}$
used here is the metric that arises naturally in
the $\sigma$-model, and is related to the Einstein metric
$G^{(E)}_{\mu\nu}$ through the relation:
$$
G^{(E)}_{\mu\nu} = e^{-\Phi} G_{\mu\nu}
\eqno\ethreenew
$$
After this field redefinition (together with a rescaling $\Phi\to 2\Phi$,
$A_\mu\to 2\sqrt 2 A_\mu$) one can recover the action of ref.\HOROWITZ,
except for the $H_{\mu\nu\rho}H^{\mu\nu\rho}$ term appearing in eq.\eone.

\item
{4.} We have truncated the action to contain only those terms that contain
two
or less number of derivatives.
Thus, for example, the Lorentz Chern Simons term has not been included in
the definition of $H_{\mu\nu\rho}$, since the corresponding terms in the
action will contain more than two derivatives.

We shall look for solutions that are independent of the time coordinate
$t$.
In the following analysis we shall use matrix notation to describe the
various fields.
In this notation, $G_{\mu\nu}$ and $B_{\mu\nu}$ will be treated as
$4\times 4$ matrices, and $A_\mu$ will be treated as a four dimensional
column vector, with the fourth row and/or column corresponding to the time
coordinate $t$.
Let us now define the matrices $\ck$, $\eta$ and $\cm$ as,
$$
\ck_{\mu\nu} =-B_{\mu\nu} -G_{\mu\nu} -{1\over 4}A_\mu A_\nu
\eqno\efour
$$
$$
\eta_{\mu\nu} = Diag (1,1,1,-1)
\eqno\efive
$$
and
$$
\cm =\pmatrix{ (\ck^T-\eta) G^{-1}(\ck -\eta) & (\ck^T-\eta) G^{-1}(\ck
+\eta) & -(\ck^T -\eta) G^{-1} A\cr
(\ck^T+\eta) G^{-1}(\ck -\eta) & (\ck^T+\eta) G^{-1}(\ck +\eta) & -(\ck^T
+\eta) G^{-1} A\cr
-A^T G^{-1} (\ck-\eta) & -A^T G^{-1} (\ck +\eta) & A^T G^{-1} A\cr
}
\eqno\esix
$$
Here $T$ denotes transposition of a matrix.
Eq.\esix\ defines a $9\times 9$ matrix $\cm$.
The result of ref.\FAWAD\ then says that if $\{G_{\mu\nu}, B_{\mu\nu},
\Phi, A_\mu\}$ describes a time independent solution of the classical
equations of motion derived from the action given in eq.\eone, then
$\{G'_{\mu\nu}, B'_{\mu\nu}, \Phi', A'_\mu\}$ also describes a solution of
the same equations of motion, if the primed variables are related to the
unprimed ones through the relation,
$$
\cm' = \Omega \cm \Omega^T, ~~~\Phi' -\ln\det G' =\Phi -\ln\det G
\eqno\eseven
$$
where,
$$
\Omega =\pmatrix{ I_7 & & \cr & \cosh\alpha & \sinh\alpha\cr & \sinh\alpha
& \cosh\alpha \cr}
\eqno\eeight
$$
Here $I_7$ denotes a $7\times 7$ identity matrix, and $\alpha$ is an
arbitrary number.
Eqs.\eseven\ uniquely determine all the primed fields in terms of the
unprimed ones.

We now apply this transformation to the charge neutral rotating black hole
solution.
This is given by the standard Kerr solution\ABS\
$$\eqalign{
ds^2 =& -{\rho^2 +a^2\cos^2\theta -2m\rho\over \rho^2
+a^2\cos^2\theta} dt^2 +{\rho^2 +a^2\cos^2\theta\over\rho^2 +a^2 -2m\rho}
d\rho^2 + (\rho^2 +a^2\cos^2\theta) d\theta^2\cr
& + {\sin^2\theta\over \rho^2 + a^2\cos^2\theta} \{(\rho^2 +a^2)(\rho^2
+a^2\cos^2\theta) +2m\rho a^2\sin^2\theta\}d\phi^2
-{4m\rho a\sin^2\theta\over \rho^2 +a^2\cos^2\theta} dt d\phi \cr
\Phi =& 0, ~~~~ B_{\mu\nu}=0,~~~~ A_\mu=0\cr
}
\eqno\enine
$$
The transformed solution is given by,
$$\eqalign{
ds^{\prime 2}=& -{(\rho^2 +a^2\cos^2\theta -2m\rho)
(\rho^2+a^2\cos^2\theta) \over (\rho^2+a^2\cos^2\theta
+2m\rho\sinh^2{\alpha\over 2})^2} dt^2\cr
& +{\rho^2 +a^2\cos^2\theta\over \rho^2
+a^2 -2m\rho} d\rho^2 + (\rho^2 +a^2\cos^2\theta) d\theta^2\cr
&+\{(\rho^2+a^2)(\rho^2+a^2\cos^2\theta) +2m\rho a^2\sin^2\theta +4m\rho
(\rho^2+a^2) \sinh^2{\alpha\over 2}+ 4m^2\rho^2\sinh^4{\alpha\over 2}\} \cr
& \qquad \times {(\rho^2+a^2\cos^2\theta)\sin^2\theta\over (\rho^2
+a^2\cos^2\theta +2m\rho\sinh^2{\alpha\over 2})^2} d\phi^2\cr
&- {4m\rho a\cosh^2{\alpha\over 2} (\rho^2+a^2\cos^2\theta)\sin^2\theta
\over (\rho^2
+a^2\cos^2\theta +2m\rho\sinh^2{\alpha\over 2})^2} dt d\phi\cr
}
\eqno\eten
$$
$$
\Phi' =-\ln {\rho^2 +a^2\cos^2\theta +2m\rho\sinh^2{\alpha\over 2} \over
\rho^2 +a^2\cos^2\theta}
\eqno\eeleven
$$
$$
A'_\phi = -{2m\rho a\sinh\alpha\sin^2\theta\over \rho^2 +a^2\cos^2\theta
+2m\rho\sinh^2{\alpha\over 2}}
\eqno\etwelve
$$
$$
A'_t = {2m\rho\sinh\alpha\over \rho^2 +a^2\cos^2\theta
+2m\rho\sinh^2{\alpha\over 2}}
\eqno\ethirteen
$$
$$
B'_{t\phi} = {2m\rho a\sinh^2{\alpha\over 2}\sin^2\theta \over \rho^2
+a^2\cos^2\theta +2m\rho\sinh^2{\alpha\over 2}}
\eqno\efourteen
$$
The other components of $A'_\mu$ and $B'_{\mu\nu}$ vanish.
The Einstein metric $ds^{\prime 2}_E\equiv e^{-\Phi'}ds^{\prime 2}$ is given
by,
$$\eqalign{
ds^{\prime 2}_E =& -{\rho^2 +a^2\cos^2\theta -2m\rho\over \rho^2
+a^2\cos^2\theta +2m\rho\sinh^2{\alpha\over 2}} dt^2
+{\rho^2
+a^2\cos^2\theta +2m\rho\sinh^2{\alpha\over 2}\over \rho^2 +a^2 -2m\rho}
d\rho^2 \cr
& +(\rho^2
+a^2\cos^2\theta +2m\rho\sinh^2{\alpha\over 2}) d\theta^2
-{4m\rho a\cosh^2{\alpha\over 2}\sin^2\theta\over \rho^2
+a^2\cos^2\theta +2m\rho\sinh^2{\alpha\over 2}} dtd\phi\cr
& +\{(\rho^2+a^2)(\rho^2+a^2\cos^2\theta) +2m\rho a^2\sin^2\theta +4m\rho
(\rho^2+a^2) \sinh^2{\alpha\over 2}+ 4m^2\rho^2\sinh^4{\alpha\over 2}\}\cr
& \times {\sin^2\theta \over \rho^2
+a^2\cos^2\theta +2m\rho\sinh^2{\alpha\over 2}} d\phi^2\cr
}
\eqno\efifteen
$$
This metric describes a black hole solution with mass $M$, charge $Q$,
angular momentum $J$, and magnetic dipole moment $\mu$ given by,
$$
M={m\over 2} (1+\cosh\alpha),~~~~ Q={m\over\sqrt 2}\sinh\alpha, ~~~~ J=
{ma\over 2} (1+\cosh\alpha),~~~~ \mu ={1\over\sqrt 2} ma\sinh\alpha
\eqno\esixteen
$$
so that the $g$-factor\HOROWITZ\ is given by,
$$
g\equiv {2\mu M\over QJ} = 2
\eqno\eeighteen
$$

We shall now analyze various properties of this solution, and also
discuss its extremal limit.
For this purpose, it will be more convenient to express $m$, $a$ and
$\alpha$ in terms of the independent physical parameters $M$, $J$ and $Q$
by inverting the relations given in eq.\esixteen.
We get,
$$
m=M-{Q^2\over 2M}, ~~~~\sinh\alpha ={2\sqrt 2 QM\over 2M^2 -Q^2}, ~~~~
a={J\over M}
\eqno\enineteen
$$
The coordinate singularities (horizon) occur on the surfaces
$$
\rho^2 -2m\rho +a^2 =0
\eqno\etwenty
$$
which gives,
$$
\rho =m\pm \sqrt{m^2-a^2} = M-{Q^2\over 2M}\pm\sqrt{(M-{Q^2\over 2M})^2
-{J^2\over M^2}}\equiv \rho_H^\pm
\eqno\etwentyfour
$$
The area of the outer event horizon with the metric given in eq.\efifteen\
is given by,
$$
A=8\pi M (M-{{Q^2\over 2M}} +\sqrt{(M-{Q^2\over 2M})^2
-{J^2\over M^2}})
\eqno\etwentyfive
$$
{}From  eq.\etwentyfour\ we see that the horizon disappears unless,
$$
|J|\le M^2-{Q^2\over 2}
\eqno\etwentyfoura
$$
Thus the extremal limit of the black hole corresponds to
$|J|\to M-{Q^2\over 2M}$.
In this limit,
$A\to 8\pi |J|$.
Hence the event horizon remains to be of finite size in this limit, as is
expected from the general arguments of ref.\HOROWITZ.
Note the amusing result that in the extremal limit the area of the event
horizon depends only on the angular momentum $J$.
Surprisingly, this result is identical to the corresponding result for the
rotating charged black hole in a different model discussed in ref.\FROLOV.

The angular velocity $\Omega$ at the horizon is determined by demanding
that the Killing vector ${\p\over\p t} +\Omega{\p\over \p \phi}$ is null
at the horizon\ABS\HOROWITZ.
In other words,
$$
G_{tt} +2 G_{t\phi}\Omega  +G_{\phi\phi}\Omega^2 =0
\eqno\etwentyeight
$$
This gives,
$$
\Omega ={J\over 2M^2} {1\over M-{Q^2\over 2M} +\sqrt{(M-{Q^2\over 2M})^2
-{J^2\over M^2}}}
\eqno\etwentyeighta
$$
As we approach the extremal limit,
$\Omega\to {1\over 2M} sign(J)$  as long as $|J|\ne 0$.
If $J=0$, then $\Omega$ vanishes, as can be directly seen from
eq.\etwentyeighta.
It is interesting to note that in the extremal limit, $|\Omega|$ depends
only on the mass of the black hole.

Finally, the surface gravity $\kappa$ (or the Hawking temperature
$T_H=\kappa/2\pi$) is calculated at the pole as,
$$
\kappa
=\lim_{\rho\to\rho_H^+}\sqrt{g^{\rho\rho}}\p_\rho\sqrt{-g_{tt}}|_{\theta
=0}
={\sqrt{(2M^2 -Q^2)^2 -4J^2}\over 2M(2M^2 -Q^2 +\sqrt{(2M^2-Q^2)^2-4J^2})}
\eqno\etwentynine
$$
Thus in the extremal limit $\kappa\to 0$ if $J\ne 0$.
On the other hand, if $J=0$, then $\kappa= {1\over 4M}$,
in agreement with the results of refs.\BLACK.

To summarize, in this paper we have constructed a rotating charged black
hole solution in four dimensional heterotic string theory and studied its
various properties.
The extremal limit of the solution was also discussed, and, for $J\ne 0$,
was found to
have features that are qualitatively similar to the extremal rotating
black hole rather than extremal charged black hole, as was
conjectured in ref.\HOROWITZ.

\vskip .5in

\centerline{\bf References}

\vskip .5in

\item
{1.} G. Gibbons and K. Maeda, Nucl. Phys. {\bf B298}, 741 (1988);
D. Garfinkle, G. Horowitz and A. Strominger, Phys. Rev. {\bf D43},
3140 (1991);
G. Horowitz and A. Strominger, Nucl. Phys. {\bf B360}, 197 (1991);
A. Shapere, S. Trivedi and F. Wilczek, Mod. Phys. Lett. {\bf A6}, 2677
(1991).

\item
{2.} R. Adler, M. Bazin and M. Schiffer, Introduction to General Relativity,
McGraw Hill, 1975; K. Thorne, R. Price and D. Macdonald, Black Holes: The
Membrane Paradigm, Yale Univ. Press, 1986.

\item
{3.} J. Horne and G. Horowitz, preprint UCSBTH-92-11.

\item
{4.} S. Ferrara, J. Scherk and B. Zumino, Nucl. Phys. {\bf B121}, 393 (1977);
E. Cremmer, J. Scherk and S. Ferrara, Phys. Lett. {\bf B68}, 234 (1977);
{\bf B74}, 61 (1978);
E. Cremmer and J. Scherk, Nucl. Phys. {\bf B127}, 259 (1977);
E. Cremmer and B. Julia, Nucl. Phys.{\bf B159}, 141 (1979);
M. De Roo, Nucl. Phys. {\bf B255}, 515 (1985); Phys. Lett. {\bf B156},
331 (1985);
E. Bergshoef, I.G. Koh and E. Sezgin, Phys. Lett. {\bf B155}, 331 (1985);
M. De Roo and P. Wagemans, Nucl. Phys. {\bf B262}, 646 (1985);
L. Castellani, A. Ceresole, S. Ferrara, R. D'Auria, P. Fre and E. Maina,
Nucl. Phys. {\bf B268}, 317 (1986); Phys. Lett. {\bf B161}, 91 (1985);
S. Cecotti, S. Ferrara and L. Girardello, Nucl. Phys. {\bf B308},
436 (1988);
M. Duff, Nucl. Phys. {\bf B335}, 610 (1990).

\item
{5.} G. Veneziano, Phys. Lett. {\bf B265}, 287 (1991);
K. Meissner and G. Veneziano, Phys. Lett. {\bf B267}, 33 (1991); Mod.
Phys. Lett. {\bf A6}, 3397 (1991);
M. Gasperini, J. Maharana and G. Veneziano, Phys. Lett. {\bf B272},
277 (1991);
M. Gasperini and G. Veneziano, preprint CERN-TH-6321-91.

\item
{6.} A. Sen, Phys. Lett. {\bf B271}, 295 (1991); {\bf 274}, 34 (1991).

\item
{7.} S.F. Hassan and A. Sen, preprint TIFR/TH/91-40 (to appear in Nucl. Phys.
B).

\item
{8.} J. Horne, G. Horowitz and A. Steif, Phys. Rev. Lett. {\bf 68}, 568 (1992);
M. Rocek and E. Verlinde, preprint IASSNS-HEP-91-68;
P. Horava, preprint EFI-91-57;
A. Giveon and M. Rocek, preprint IASSNS-HEP-91-84;
S. Kar, S. Khastagir and A. Kumar, preprint IP-BBSR-91-51;
J. Panvel, preprint LTH 282.

\item
{9.} V. Frolov, A. Zelnikov and U. Bleyer, Ann. Phys. (Leipzig) {\bf 44}, 371
(1987).

\end